\documentclass[aps,prb,twocolumn,superscriptaddress]{revtex4-2}
\usepackage{graphicx}
\usepackage{color}
\usepackage[tight]{units}
\usepackage{epstopdf}
\usepackage[normalem]{ulem}
\DeclareGraphicsExtensions{.pdf,.jpg,.png,.eps}
\usepackage{amsfonts}
\usepackage{amsmath}
\usepackage{amssymb}
\usepackage{color}
\usepackage{color}
\usepackage{lineno}
\usepackage{epstopdf}
\usepackage{adjustbox}

\usepackage[T1]{fontenc}
\usepackage{hyperref}
\usepackage{xcolor}

\newcommand{\suraj}{\color{black}}
\newcommand{\theory}{\color{black}}

\newcommand{\s}{Co$_2$(Ti$_{0.25}$V$_{0.25}$Cr$_{0.25}$Fe$_{0.25}$)Al}

\begin{document}

\title{ Cocktail effect and robust Berry curvature driven anomalous Hall conductivity in the entropy-stabilized Heusler alloy \s}

\author{Suraj Kushwaha}
\affiliation{Department of Physics and Material Science, Thapar Institute of Engineering and Technology, Punjab 147004, India}

\author{S. K. Panda}
\affiliation{Department of Physics, Bennett University, Greater Noida, Uttar Pradesh 201310, India}

\author{Sourav Marik}
\affiliation{Department of Physics and Material Science, Thapar Institute of Engineering and Technology, Punjab 147004, India}

\author{Kartik Samanta}
\affiliation{Department of Physics and Materials Science and Engineering (PMSE),
Jaypee Institute of Information Technology (JIIT), Wish Town Campus, Noida, UP-201309, India}

\author{Tirthankar Chakraborty}
\email{tirthankar@thapar.edu}
\affiliation{Department of Physics and Material Science, Thapar Institute of Engineering and Technology, Punjab 147004, India}


\begin{abstract}
The interplay between chemical disorder and persistence of Berry curvature driven transport phenomena remains an important open question in entropy-stabilized systems. Here, we synthesize an entropy-stabilized Heusler alloy \s~ and systematically investigate its structural, magnetic, and magnetotransport properties using a combination of experimental measurements and density functional theory (DFT) calculations. The system crystallizes in cubic space group $Fm\Bar{3}m$ and exhibits ferromagnetism with saturation magnetization in close agreement with the Slater--Pauling prediction. Transport and magnetotransport measurements reveal metallic behavior and a pronounced anomalous Hall effect with an anomalous Hall conductivity of approximately $134.4~ \Omega^{-1}$.cm$^{-1}$. 
Combined experimental observations and first-principles calculations establish that the anomalous Hall effect is predominantly intrinsic in origin and originates from the Berry curvature of the electronic bands.
Remarkably, despite the substantial configurational disorder and the dilution of the constituent parent compounds, the anomalous Hall conductivity remains comparable to the largest values reported in the corresponding parent Heusler systems. This behavior reflects the manifestation of the cocktail effect, one of the core characteristics of entropy-stabilized systems. Our results also demonstrate that Berry curvature mediated transport persists in this chemically disordered system and indicates that entropy engineering can be a promising route for tuning intrinsic anomalous Hall responses.
\end{abstract}
\keywords{Keywords}

\maketitle

\section{Introduction}
Heusler alloys have emerged as an important class of quantum materials owing to their rich and novel structural, magnetic, electronic and thermodynamic properties \cite{felser2015heusler,hirohata2022heusler, al2025physical, tavares2023heusler, chen2024thermoelectric, quinn2021advances, wederni2024crystal}. In particular, Co-based full Heusler alloys with the general formula Co$_2$YZ have attracted sustained attention because of their tunable magnetic properties, high Curie temperatures, half-metallicity, and anomalous transport phenomena often originating from nontrivial band topology \cite{co2mnal_nature}. Their highly flexible electronic structure, arising from strong hybridization between transition metal 3d states, makes them promising candidates for spintronic and topological transport applications.

Among the various transport phenomena observed in magnetic Heusler systems, the anomalous Hall effect (AHE) has received considerable interest as an effective probe of the underlying electronic band topology. Modern developments have established that the intrinsic contribution to AHE originates from the Berry curvature of occupied electronic bands in momentum space \cite{karplus1954hall, nagaosa2010anomalous, onoda2006intrinsic}

In recent years, several Heusler compounds have been shown to exhibit predominantly intrinsic AHE associated with Berry-curvature-driven transport \cite{sakuraba2020giant,wederni2024crystal, manna2018colossal, co2mnal_nature,shahi2022antisite}. Such systems therefore provide a fertile platform for exploring the Berry curvature driven phenomena. However, most of such investigations in Heusler alloys have been focused on chemically ordered compounds so far. 
 
Recently, entropy engineering has emerged as an effective strategy for stabilizing materials. Resulting systems are called the high entropy materials, which consist of five or more elements mixed in near-equiatomic proportions \cite{yeh2004nanostructured, george2019high}. The formation of single-phase solid solutions in these systems is thermodynamically favored by the large configurational entropy of mixing ($\Delta S_{\mathrm{mix}}$), which contributes significantly to lowering the Gibbs free energy. The resulting atomic-scale disorder, arising from the random occupation of lattice sites by multiple elements with different atomic sizes and electronic configurations, has been associated with a variety of unusual physical properties and exceptional mechanical performance \cite{ye2016high}.
In Heusler systems, multi-element substitution provides an additional degree of freedom for tuning the electronic structure and magnetic exchange interactions. The resulting chemical disorder can significantly influence the electronic and transport properties, making the understanding of disorder effects increasingly important in functional Heusler materials \cite{sharma2025multiband,tian2009proper,hou2015multivariable,kudrnovsky2013anomalous,Co2FeAl_shukla,vilanova2011influence,luo2016competition}. More recently, the role of configurational entropy and severe chemical disorder has attracted considerable attention owing to its profound impact on the structural, magnetic, electronic, and superconducting properties of a wide range of materials \cite{sharma2025multiband,jangid2024superconductivity,rajeswari2025large}. Despite this growing interest, experimental investigations of entropy-stabilized systems remain relatively limited. Consequently, a fundamental open question arises whether intrinsic Berry curvature mediated transport can survive in entropy-stabilized systems where chemical disorder is inherently strong and and how such configurational entropy influences the associated phenomena.

In this work, we investigate the anomalous Hall effect in the entropy-stabilized Heusler alloy \s~through a combination of structural, magnetic, magnetotransport, and first-principles calculations. X-ray diffraction confirms the formation of a cubic Heusler phase with signatures of chemical disorder, while magnetization measurements reveal robust ferromagnetism with a saturation moment in excellent agreement with the Slater--Pauling prediction. The alloy exhibits exhibits metallic transport and a pronounced anomalous Hall effect with a maximum anomalous Hall conductivity of approximately 134.4 $\Omega^{-1}$.cm$^{-1}$. With combined experimental and theoretical analyses, we establish that the AHE is predominantly intrinsic in origin and driven by the Berry curvature of the electronic bands. Remarkably, despite substantial configurational disorder and extensive chemical dilution of the constituent parent compounds, the anomalous Hall conductivity remains comparable to the highest values reported among the corresponding parent Heusler systems. Our results reflect manifestation of the cocktail effect and demonstrate that Berry-curvature-mediated transport remains remarkably robust against chemical disorder.
    
\section{Experimental Details}
\subsection{ Synthesis methods}

Entropy stabilized Heusler alloy \s~ was synthesized in polycrystalline form using the vacuum arc melting technique. The constituent elements, cobalt pieces (purity 99.99\%), titanium pieces (purity 99.9\%), vanadium pieces (purity 99.9\%), chromium pieces (purity 99.9\%), iron pieces (purity 99.99\%), and aluminium pieces (purity 99.99\%) were melted in a Ti-gettered ultrahigh-purity Ar atmosphere. The alloy was melted at high temperature and subsequently flipped and remelted several times to ensure chemical homogeneity and compositional uniformity throughout the ingot. The weight loss after melting was found to be negligible.
\subsection{Structural and microstructural characterizations}
The structural characterization of the synthesized sample was carried out using a Rigaku x-ray diffractometer with Cu K$\alpha$ radiation ($\lambda = 1.5406$ \AA). The microstructural analysis was performed using a ZEISS GEMINI field-emission scanning electron microscope (FESEM), while elemental composition and stoichiometry were investigated through energy dispersive x-ray spectroscopy (EDS) using a BRUKER XFlash 6160 system. Vickers microhardness measurements were performed on the polished surfaces using an OMNITECH MVH-1C microhardness tester with an applied load of 500 g and a dwell time of 20 s.
\subsection{X-ray photoelectron spectroscopy measurements}
X-ray photoelectron spectroscopy (XPS) and ultraviolet photoelectron spectroscopy (UPS) measurements were carried out on the bulk polycrystalline \s~ sample using a Thermo Scientific Nexsa G2 series system with a spot size of $400~\mu\mathrm{m}$.
\subsection{Magnetic and physical property measurements}
Magnetic measurements were carried out using a vibrating sample magnetometer (VSM) attached to the Quantum Design Physical Property Measurement System (QD-PPMS). Magnetic hysteresis ($M$–$H$) measurements were performed at different temperatures to investigate the magnetic response of the sample.

The longitudinal resistivity and magnetotransport measurements were carried out using the  QD-PPMS over a wide temperature and magnetic field range.
{\theory
\subsection{DFT calculations}
The electronic structure calculations were performed based on density functional theory (DFT) using the plane-wave projected augmented wave (PAW) method as implemented in the Vienna ab initio Simulation Package (VASP) \cite{kresse1996efficient, kresse1999ultrasoft, blochl1994projector}. For the self-consistent calculations, we used a 6$\times$6$\times$6 k-points mesh. This choice of the k-mesh and a plane-wave cutoff of 500eV were found to provide a good convergence of the total energy. We used the Perdew-Burke-Ernzerhof (PBE) \cite{perdew1996generalized} exchange correlation functional within the generalized gradient approximation (GGA). To model the \s, we employed the Virtual Crystal Approximation (VCA)\cite{vca} as implemented in the VASP code to describe the mixed transition-metal occupancy at the B-site. The orbital-projected electronic band structures were analyzed and visualized using sumo\cite{ganose2018sumo}.

The intrinsic anomalous Hall conductivity (AHC) was evaluated using the Berry-curvature formalism within the framework of Wannier interpolation \cite{wang2006ab}. First, maximally localized Wannier functions (MLWFs) were constructed from the spin-polarized GGA+SOC Bloch states using the Wannier90 package \cite{th8_marzari2012maximally, th9_souza2001maximally, th10_mostofi2014updated}. Atomic-like Co-d, Fe-d,Ti-d,V-d, Cr-d and Al-p orbitals were used as the initial projections to generate an accurate tight-binding Hamiltonian, reproducing the first-principles electronic structure within an energy window of approximately $\pm$1 eV around the Fermi level.
With the tight-binding model Hamiltonian, we calculated the intrinsic AHC using the linear response Kubo formula approach as follows \cite{th11_yao2004first}:
\begin{equation}
			\begin{aligned}
			\Omega_{n,ij}(\vec{k}) = Im \sum_{m \neq n} \frac{ \langle n(\vec{k})| \hat{ v}_{i}|m(\vec{k})\rangle \langle m(\vec{k})|\hat{v}_{j}|n(\vec{k})\rangle-(i\leftrightarrow j)}{(\epsilon_{n}-\epsilon_{m})^{2}}
\end{aligned}
			\label{eq-BC}
\end{equation}
		where, $\Omega_{n,ij}(\vec{k})$ is the Berry curvature of band $n$, $\langle m(\vec{k})|$ and $\langle n(\vec{k})|$ are the eigenstates, $\epsilon$ are the eigen energies of the Hamiltonian $H$, and $\hat{v}_i$ is the velocity operator.
        Subsequently, we calculated the AHC, given by:
        \begin{equation}
			\begin{aligned}
				\sigma^{A}_{H}=& \frac{-e^{2}}{\hbar} \sum_{n} \int_{BZ} \frac{d{\vec k}}{\left(2 \pi\right)^3}\Omega_{n,ij}(\vec{k}),
			\end{aligned}
			\label{theory_sigmaH}
		\end{equation}
		We used a k-point mesh of $300 \times 300 \times 300$ for the calculation of the AHC using  Eq.\ref{theory_sigmaH}.
}
\section{Results and discussions}
 \subsection{Structural characterizations}
\begin{figure*}
\centering
\includegraphics[width=\textwidth]{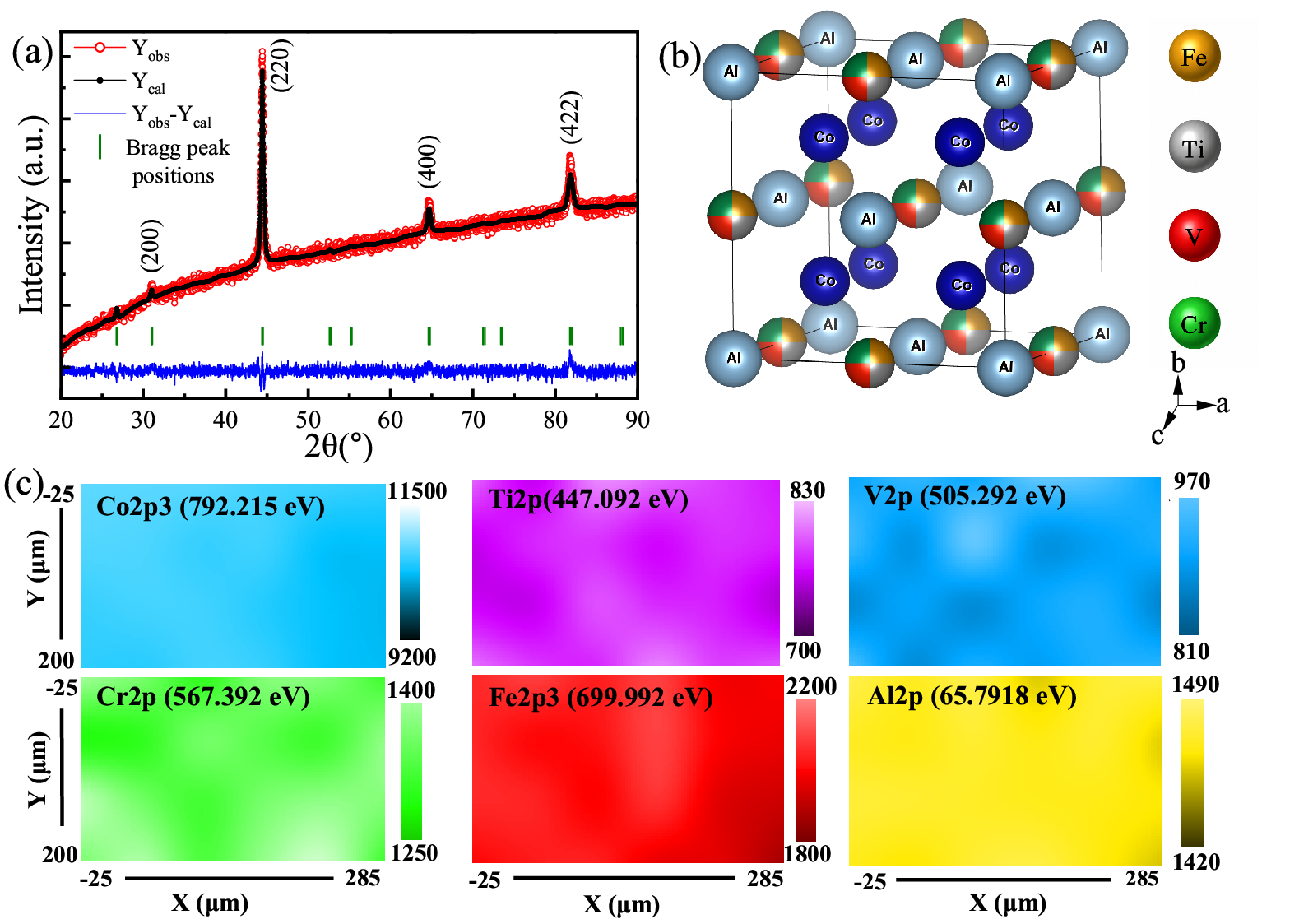}
\caption{(a) Room-temperature x-ray diffraction pattern along with Le Bail refinement, (b) crystal structure and (c) XPS elemental area mappings illustrating the spatial distribution of Co, Ti, V, Cr, Fe, and Al of the entropy stbilized system \s. No significant large-scale elemental segregation is observed within the scanned region.
}.
\label{fig:XRD_XPS}
\end{figure*}

 The room-temperature powder X-ray diffraction pattern and corresponding Le Bail refinement of the entropy stabilized Heusler alloy \s~ are shown in Fig.~\ref{fig:XRD_XPS}(a). Significant broadening of the peaks can be observed. This feature can be observed in many other high entropy alloys and can be attributed to the disorder caused by different atomic radii \cite{sharma2025superconductivity,     rajeswari2025large} [and other few papers of high entropy-our group]. Furthermore, the system appears to be hard and brittle and therefore it is difficult to ensure fine grain size when it is powdered. In addition, due to similar scattering factors of the atoms, it is not ideal to analyze the X-ray diffraction pattern by Rietveld refinement. Instead, profile fitting with Le Bail fit is more appropriate. The system is found to crystallize in space group $Fm\overline{3}m$ (225). 
 It is worth mentioning here that the $L2_1$  ordered cubic structure with space group $Fm\overline{3}m$ and the atomic Wyckoff positions with Co at $8c$ (0.25, 0.25, 0.25), Ti/V/Cr/Fe at $4b$ (0.5, 0.5, 0.5) and Al at $4a$ (0, 0, 0) can have disorders which lead to the $B2$ or $A2$ type structures \cite{co2mnal_nature, luo2016competition, vilanova2011influence, sakuraba2020giant}. In the present system, absence of the (1 1 1) but presence of the (2 0 0) reflection indicates the $B2$ type structure where there is antisite disorder between the Ti/V/Cr/Fe at $4b$ and Al at $4a$ sites \cite{mdpi_structure_review, pathak2025significance, lakhni2022antisiteXA, bhowmik2026disproportionate}. This is also consistent with the more electronegative nature of Co and therefore likely to be occupying $8c$. The profile fitting matches nicely with the experimental pattern considering the $B2$ structure with lattice parameter 5.765~\AA ,which is very similar to the corresponding parent Heusler compositions \cite{Co2TiAl_intrinsic, Co2TiAl_extrinsic, Co2VAl_AHE_intrinsic, Co2VAl_1, Co2CrAl_AHE_temperaure_dependent, Co2CrAl_AHE_theory, Co2FeAl_shukla}.
 
 The corresponding crystal structure is illustrated in Fig.~\ref{fig:XRD_XPS}(b). XPS elemental area mappings Fig.~\ref{fig:XRD_XPS}(c) reveal the spatial distribution of Co, Ti, V, Cr, Fe, and Al across the investigated region. No significant element-rich regions or large-scale chemical segregation are observed which indicates a reasonably uniform distribution of the constituent elements and supports the successful formation of the entropy-stabilized Heusler alloy. 

\subsection{Microhardness measurements}

{\suraj A Vickers microhardness of $\sim$582.9 HV ($\sim$5.71 GPa) is observed for \s~, which is considerably higher than several reported FCC high-entropy alloys and comparable to hard Co$_2$Fe-based Heusler systems \cite{fu2016microstructure,wu2006adhesive}. The enhanced hardness indicates improved resistance against plastic deformation and suggests good mechanical stability of the investigated alloy. Previous studies have shown that atomic ordering, phase constitution, and elemental substitution strongly influence hardness behaviour in Heusler and high-entropy alloy systems \cite{zheng2019influence,wu2006adhesive}. }
\subsection{X-ray photoelectron spectroscopy(XPS)}
{\suraj 
In order to investigate the electronic structure and chemical states of the constituent elements in \s, X-ray photoelectron spectroscopy (XPS) and ultraviolet photoelectron spectroscopy (UPS) measurements were performed. The XPS measurements were carried out using Al K$\alpha$ radiation ( 1486.6 eV), while He I radiation ( 21.22 eV) was employed for UPS measurements. Prior to the photoemission measurements, the sample surface was etched using Ar$^{+}$ ions with an energy of 3000 eV for 200 s in order to remove the surface oxide layer and possible contaminants. XPS was used to probe the core-level binding energies and chemical states of the constituent elements}

\begin{figure*}
\centering
\includegraphics[width=0.75\textheight]{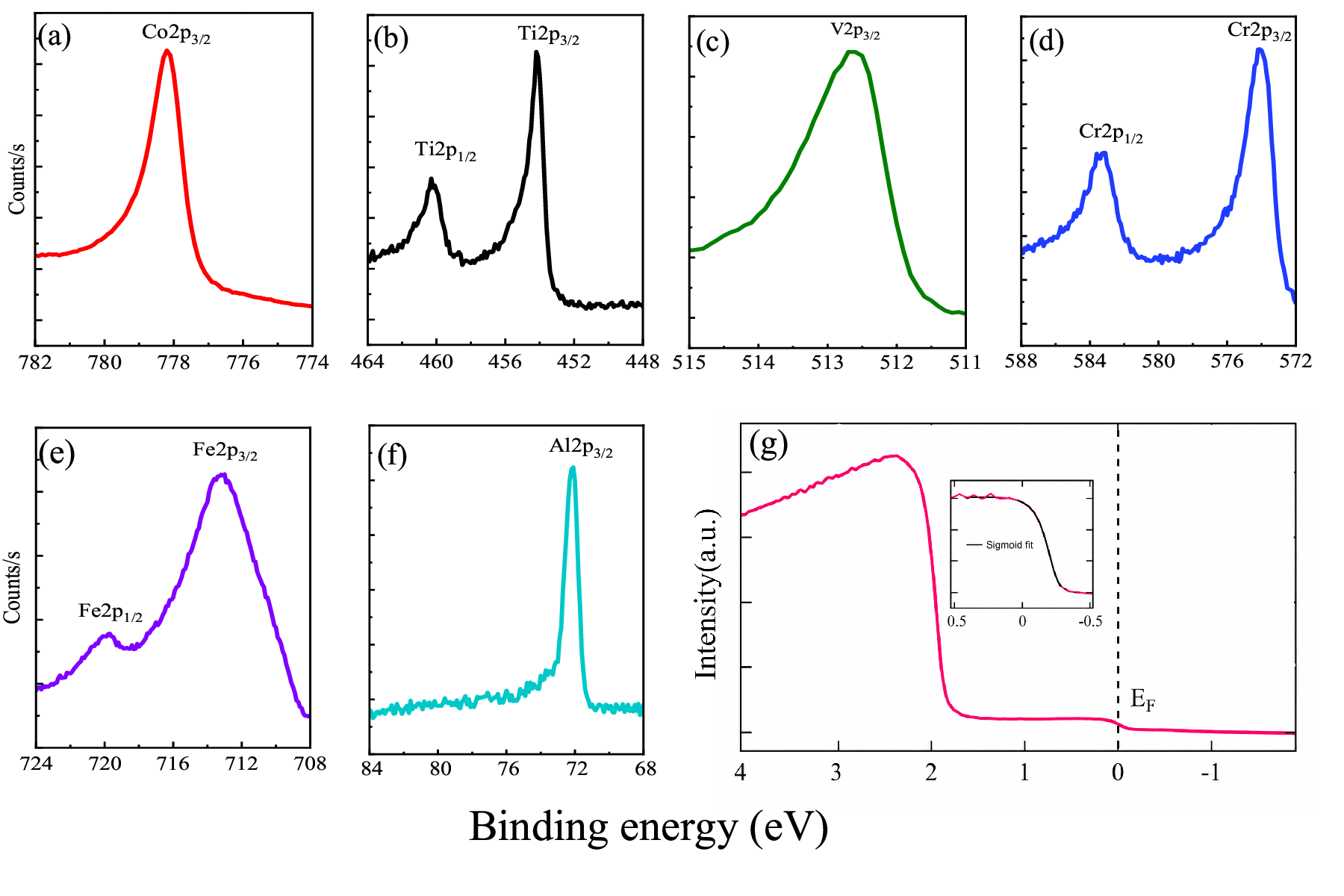}
\caption{(a--f) High-resolution XPS spectra of Co 2$p$, Ti 2$p$, V 2$p$, Cr 2$p$, Fe 2$p$, and Al 2$p$, respectively, for \s. The observed core-level features confirm the incorporation of all constituent elements within the Heusler alloy. (g) UPS spectrum showing the electronic states near the Fermi level along with the Sigmoid fit shown in inset.}
\label{fig:XPS_Area_Mapping_UPS}
\end{figure*}

{\suraj The high-resolution core-level spectra of Co, Ti, V, Cr, Fe, and Al are shown in Fig.~\ref{fig:XPS_Area_Mapping_UPS}(a--f). The Co 2$p_{3/2}$ peak located near $\sim$778 eV is characteristic of metallic Co in Heusler intermetallics, while the absence of pronounced satellite features indicates negligible surface oxidation \cite{kumar2019x}. The Fe 2$p$ spectrum exhibits well-defined 2$p_{3/2}$ and 2$p_{1/2}$ spin--orbit split components, confirming the metallic bonding environment of Fe \cite{yamashita2008analysis}. Similarly, the Ti 2$p$, V 2$p$, and Cr 2$p$ spectra display characteristic core-level features consistent with their incorporation into the alloy lattice \cite{biesinger2010resolving,biesinger2011resolving,moulder1992handbook}. The Al 2$p$ peak further confirms the presence of Al within the Heusler framework.


 Figure \ref{fig:XPS_Area_Mapping_UPS} (g)   shows the He~I valence-band UPS spectrum of the alloy. The occupied spectral weight near the Fermi level was evaluated from the normalized intensity in the 0--1~eV binding-energy range, which indicates the states closest to $E_F$ \cite{whitten2023ultraviolet}. The spectrum shows finite intensity at $E_F$. The extracted spectral weight near-$E_F$ is approximately 0.92. However,  broader 0--4~eV range is shown to visualize the overall valence-band structure. The inset of Figure \ref{fig:XPS_Area_Mapping_UPS} (g) shows the sigmoid fit used to determine the Fermi-edge position \cite{hufner2013photoelectron} of the spectra. The nonzero spectral weight at $E_F$ confirms metallic character of the system.
 }

 \subsection{Magnetic properties}
The isothermal magnetization ($M$--$H$) curves measured at different temperatures reveal rapid magnetic saturation at a relatively low applied field of approximately 2000 Oe as shown in Fig.~\ref{fig:AHE}(a). An enlarged view of the low-field region is represented in the inset of Fig.~\ref{fig:AHE}(a), from which the coercive field at 300 K is estimated to be about 17 Oe, confirming the soft ferromagnetic nature of the investigated alloy. The saturation magnetization ($M_s$) is found to be about $2.9~\mu_B$/f.u. at 5 K and exhibits only a slight reduction with increasing temperature, reaching nearly $2.5~\mu_B$/f.u. at 300 K.
The weak temperature dependence of $M_s$ suggests robust ferromagnetic ordering persisting up to room temperature. Furthermore, the experimentally obtained saturation magnetic moment agrees well with the value of $2.8~\mu_B$/f.u. predicted from the Slater--Pauling rule, indicating consistency between the experimental observations and the expected electronic structure of the Heusler alloy.

 \subsection{Transport and anomalous Hall}
The temperature dependence of the longitudinal resistivity $\rho_{xx}$ of \s~ exhibits an overall metallic behavior over the measured temperature range. The residual resistivity ratio (RRR) is estimated to be approximately 1.14, indicating the presence of significant disorder and enhanced defect-induced scattering in the system. Such a relatively low RRR is commonly observed in chemically disordered and entropy-stabilized alloys, where atomic-scale randomness strongly influences charge transport. The obtained value is comparable to those reported for analogous parent Heusler systems such as Co$_2$VAl and Co$_2$FeAl \cite{Co2VAl_AHE_intrinsic, Co2FeAl_shukla}, as well as several recently reported high-entropy and medium-entropy alloys \cite{sharma2025multiband, egilmez2021magnetic, motla2023superconducting, jangid2024superconductivity}.

In metallic systems, the electrical resistivity generally originates from multiple scattering mechanisms, primarily associated with structural defects and phonons. According to Matthiessen’s rule, the total resistivity can therefore be expressed as
\begin{align}
\rho_{xx}(T)=\rho_0+\rho_{ph}
\label{BG}
\end{align}
where $\rho_0$ represents the temperature-independent residual resistivity arising from static disorder and defects, while $\rho_{ph}$ corresponds to the phonon scattering contribution. The phonon contribution to $\rho_{xx}(T)$ can be described using the Bloch--Grüneisen (BG) formalism \cite{grimvall1976electron}
\begin{align}
\rho_{ph}(T)=A \left(\frac{T}{\theta_D}\right)^5 \int_{0}^{\theta_D/T} \frac{x^5}{(e^x-1)(1-e^x)} ,dx
\label{BG-elph-quation}
\end{align}
where $A$ denotes the electron--phonon coupling strength and $\theta_D$ is the Debye temperature.
We performed the BG fit on the temperature dependent resistivity data using the contributions described in Eq.~\ref{BG}. The fitting yields $\rho_0=208.69~\mu\Omega$.cm, $\theta_D=105.59~K$, and $A=37.73~\mu\Omega$.cm. The obtained Debye temperature is comparable to those reported for several entropy-stabilized alloy systems \cite{rajeswari2025large, sharma2025multiband}.
In conventional ferromagnets, the magnon scattering contribution is typically incorporated by adding a term $\alpha T^2$ to Eq.~\ref{BG}. In the present case, however, the resistivity data can be satisfactorily described without including any magnon contribution. Nevertheless, fitting attempts incorporating the magnon scattering term resulted in an extremely small coefficient of the order of $\sim10^{-12}~\Omega$.cm K$^{-2}$, indicating that the magnon contribution to the electrical resistivity is negligible within the investigated temperature range. Therefore, the temperature dependence of $\rho_{xx}$ is predominantly governed by electron--phonon scattering processes.
 
\begin{figure*}
\centering
\includegraphics[width=0.75\textheight]{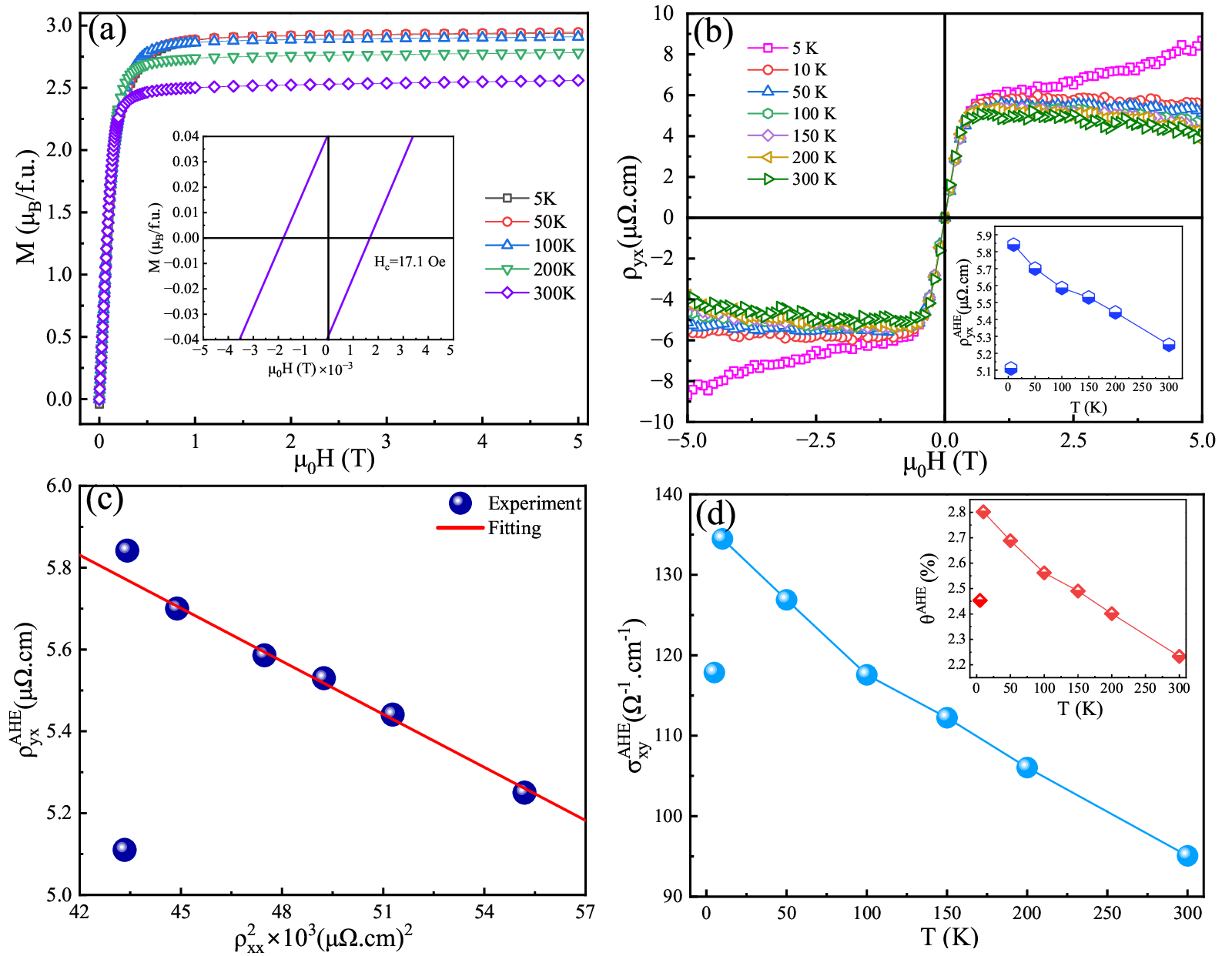}
\caption{\label{fig:AHE}
(a) Isothermal magnetization curves of \s~ measured at different temperatures. The inset shows an enlarged view of the low-field region at 300 K. (b) Field dependence of the Hall resistivity, $\rho_{yx}$, measured at various temperatures. The inset displays the temperature dependence of the anomalous Hall resistivity, $\rho_{yx}^{\mathrm{AHE}}$. (c) Scaling behavior of $\rho_{yx}^{\mathrm{AHE}}$ as a function of $\rho_{xx}^{2}$. The linear dependence suggests that the anomalous Hall effect is predominantly governed by the intrinsic mechanism. (d) Temperature dependence of the anomalous Hall conductivity, $\sigma_{xy}^{\mathrm{AHE}}$. The inset shows the variation of the anomalous Hall angle, $\theta^{\mathrm{AHE}}$, with temperature.}
\end{figure*}

In order to gain deeper insight into the electronic transport properties, we investigated the transverse resistivity $\rho_{yx}$ of the material as a function of magnetic field and temperature.  Fig.~\ref{fig:AHE}(b) presents the field dependence of the Hall resistivity measured at different temperatures. The Hall signal exhibits a sharp increase in the low-field region followed by a nearly linear variation at high fields, indicating the presence of anomalous Hall effect in the system. In magnetic metals, the total Hall resistivity $\rho_{yx}$ generally consists of both ordinary and anomalous contributions and can be expressed empirically as \cite{karplus1954hall}
\begin{align}
\rho_{yx}=\rho_{yx}^O (B)+\rho_{yx}^{AHE} (M)
\label{eqn-rhoxy-ordinary-anomalous}\\
=R_0\mu_0H+R_SM
\label{eqn-rhoxy-ordinary-anomalous-coefficient}
\end{align}
where $B=\mu_0H$ is the magnetic induction, and $\rho_{yx}^O$ and $\rho_{yx}^{AHE}$ correspond to the ordinary and anomalous Hall resistivities, respectively. Here, $R_0$ and $R_S$ denote the ordinary and anomalous Hall coefficients, while $\mu_0$ is the permeability of free space. The ordinary Hall coefficient ($R_H$) associated with $\rho_{yx}^O$ can be determined from the high field slope $d\rho_{yx}/dH$, where the magnetization is already saturated. The anomalous Hall resistivity $\rho_{yx}^{AHE}$ can be obtained by extrapolating the high field linear region of the $\rho_{yx}$ curve to zero magnetic field. The extracted $\rho_{yx}^{AHE}$ and its variation with temperature is shown in inset of Fig.~\ref{fig:AHE}(b).

Interestingly, the slope of $\rho_{yx}$ in the high-field region is positive at 5 K, whereas it becomes weakly negative at 10 K and above. This behavior suggests that hole-like carriers dominate the electrical transport at 5 K, whereas electron-like carriers become dominant at higher temperatures. Such a sign reversal of the ordinary Hall coefficient may indicate the presence of multiband transport and possible carrier compensation effects in the present entropy-stabilized system. The charge carrier density $n$, estimated using the relation $R_0=\frac{1}{ne}$, is found to be of the order of $2.3\times 10^{20}$ cm$^{-3}$, which is consistent with metallic transport in moderately disordered Heusler alloys \cite{Co2TiAl_intrinsic, Co2TiAl_extrinsic, Co2VAl_AHE_intrinsic, Co2VAl_1, Co2CrAl_AHE_temperaure_dependent, Co2CrAl_AHE_theory, Co2FeAl_shukla}. 
In general, the anomalous Hall effect (AHE) arises from both intrinsic and extrinsic mechanisms. The intrinsic contribution, originally proposed by Karplus and Luttinger (KL) \cite{karplus1954hall}, originates from the combined effect of spin--orbit coupling (SOC) and broken time-reversal symmetry in the electronic band structure of a ferromagnetic material. Subsequent theoretical developments established that the KL mechanism can be understood in terms of the Berry curvature of Bloch electrons in momentum space \cite{nagaosa2010anomalous}. In contrast, the extrinsic contribution to AHE originates from scattering processes experienced by conduction electrons in the presence of impurities, disorder, and other imperfections in the crystal potential. These extrinsic mechanisms are commonly classified into skew scattering and side-jump scattering contributions.
From a fundamental perspective, the intrinsic origin of AHE is of particular interest, as it is directly linked to the electronic band structure and Berry curvature effects, and is therefore highly relevant for potential spintronic and topological transport applications. To further analyze the relative contributions of different scattering processes to the AHE, we employ the scaling relation proposed by Tian \textit{et al.} \cite{tian2009proper, hou2015multivariable, wu2016anomalous}, commonly referred to as the TYJ model, expressed as
\begin{align}
\rho_{yx}^{AHE}=\alpha \rho_{xx_0} + \beta \rho_{xx_0}^2 + \gamma \rho_{xx}^2
\label{scaling-equation}
\end{align}
where $\rho_{xx_0}$ denotes the residual resistivity. Here, the linear term is associated with the extrinsic skew-scattering mechanism, the quadratic residual resistivity term describes the side-jump contribution, while the term proportional to $\rho_{xx}^2$ represents the intrinsic anomalous Hall contribution arising from the Berry curvature of the electronic bands. Here, $\alpha$, $\beta$, and $\gamma$ are the coefficients corresponding to skew scattering, side-jump scattering, and intrinsic mechanisms, respectively. In the present system, the longitudinal conductivity $\sigma_{xx}$ is of the order of $10^3~\Omega^{-1}$.cm$^{-1}$, indicating that the material does not lie in the extremely clean transport regime where skew scattering typically dominates. Therefore, the contribution from skew scattering is expected to be negligible, and the scaling behavior can be primarily analyzed within the regime of $\rho_{yx}^{AHE}\propto \rho_{xx}^2$, which reflects the dominance of intrinsic and/or side-jump mechanisms. 
{Fig. \ref{fig:AHE}(c)} presents the variation of $\rho_{yx}^{AHE}$ as a function of $\rho_{xx}^2$. It follows linear behavior from 10 K up to 300 K, indicating that the observed AHE is predominantly governed by intrinsic and/or side-jump mechanisms.
The side-jump contribution to the anomalous Hall conductivity can be approximately estimated using the relation $\frac{e^2}{ha}\left(\frac{\epsilon_{SO}}{E_F}\right)$, where $\epsilon_{SO}$ and $E_F$ represent the spin--orbit interaction energy and Fermi energy, respectively \cite{onoda2006intrinsic, nozieres1973simple}. Here, $e$, $h$, and $a$ denote the electronic charge, Planck’s constant, and lattice parameter. For most ferromagnetic metals, the ratio $\epsilon_{SO}/E_F$ is typically of the order of $10^{-2}$, indicating that the side-jump contribution to the anomalous Hall conductivity should be much smaller than the intrinsic contribution. Therefore, the scaling behavior observed in Fig.~\ref{fig:AHE}(c) suggests that the AHE in the present system is predominantly intrinsic in nature. 

The anomalous Hall conductivity is calculated using the tensor conversion $\sigma_{xy}^{AHE}=\frac{\rho_{yx}^{AHE}}{\left(\rho_{yx}^{AHE}\right)^2+\left(\rho_{xx}\right)^2}$ \cite{co2mnal_nature, manna2018colossal}. The temperature dependence of $\sigma_{xy}^{AHE}$ is shown in  Fig.~\ref{fig:AHE}(d). The value of $\sigma_{xy}^{AHE}$ reaches a maximum of $134.4~ \Omega^{-1}$.cm$^{-1}$ at 10 K and gradually decreases to $95.0~\Omega^{-1}$.cm$^{-1}$ at 300 K. 

It is observed that the magnitude of $\sigma_{xy}^{AHE}$ in the present system is comparable to those reported for the corresponding parent Co-based Heusler compounds. In particular, the obtained value is larger than those reported for Co$_2$VAl and Co$_2$CrAl, while remaining somewhat lower than Co$_2$FeAl \cite{Co2VAl_AHE_intrinsic, Co2VAl_1, Co2CrAl_AHE_temperaure_dependent, Co2CrAl_AHE_theory, Co2FeAl_shukla}. In contrast, the AHE in Co$_2$TiAl has been reported to be predominantly governed by extrinsic skew-scattering rather than intrinsic mechanisms \cite{Co2TiAl_extrinsic}. The observation of a descent anomalous Hall conductivity despite the presence of considerable configurational disorder suggests that the Berry-curvature-mediated intrinsic contribution remains robust in the entropy-stabilized system. 

In addition to the anomalous Hall conductivity, the anomalous Hall angle (AHA) is another important quantity which is defined as $\theta^{AHE}=\frac{\sigma_{xy}^{AHE}}{\sigma_{xx}}$. It represents the efficiency of the transverse Hall current generated with respect to the longitudinal charge current. The temperature evolution of $\theta^{AHE}$ is displayed in the inset of  Fig.~\ref{fig:AHE}(d), where a systematic decrease is observed with increasing temperature. Notably the AHA is considerably higher than that of the parent systems as well as other high entropy systems \cite{Co2VAl_AHE_intrinsic, Co2VAl_1, Co2CrAl_AHE_temperaure_dependent, Co2CrAl_AHE_theory, Co2FeAl_shukla, rajeswari2025large}. A larger AHA reflects a stronger transverse charge response and is therefore suitability of the present entropy-stabilized system for spintronic applications. 
{\theory
\subsection{First-principles results}
We further performed first-principles density functional theory (DFT) calculations on the \s~ system to obtain information about the electronic structure and the intrinsic contribution of the Berry curvature to the anomalous Hall conductivity (AHC). We considered the experimentally measured cubic structures for the DFT calculations. The total magnetic moment per formula unit (f.u.) is found to be around 2.75 $\mu_B/f.u.$, which is in good agreement with the experimentally measured value.
In Fig. \ref{fig-band-structure} (left pannel), shown in green color, we present the electronic band structures of the system, along the high-symmetry path $\Gamma-X-W-K-\Gamma-L-U-W-L-K$ of the cubic Brillouin zone (BZ). The calculations were performed within the GGA+SOC framework considering the spins fully polarized along the c-axis. We find a metallic ground state. 
Next, we computed the intrinsic contribution to the AHC by integrating the Berry curvature over the entire BZ. Using the tight-binding Hamiltonian constructed from maximally localized Wannier functions, we calculated the AHC as a function of the Fermi energy which is shown in right pannel of Fig. \ref{fig-band-structure}. At the Fermi energy ($E_F$), the AHC is found to be $-301.9~ \Omega^{-1}$.cm$^{-1}$ for the \s~ system. It is noteworthy that the calculated value is of the same order of magnitude as the experimentally measured AHC, although the latter is somewhat smaller. Such quantitative differences may arise due to the presence of chemical disorder, finite-temperature effects, and grain-boundary scattering in the polycrystalline sample.
To understand the large contribution to the AHC around the Fermi energy in the system, we further calculated the Berry curvature contributions along the high-symmetry lines of the cubic Brillouin zone which is shown in left pannel of Fig. \ref{fig-band-structure}. We find a large redistribution of the Berry curvature along the high-symmetry path and significantly enhanced Berry-curvature contributions in the high-entropy alloy.
\begin{figure*}
\centering
\includegraphics[width=0.7\textheight]{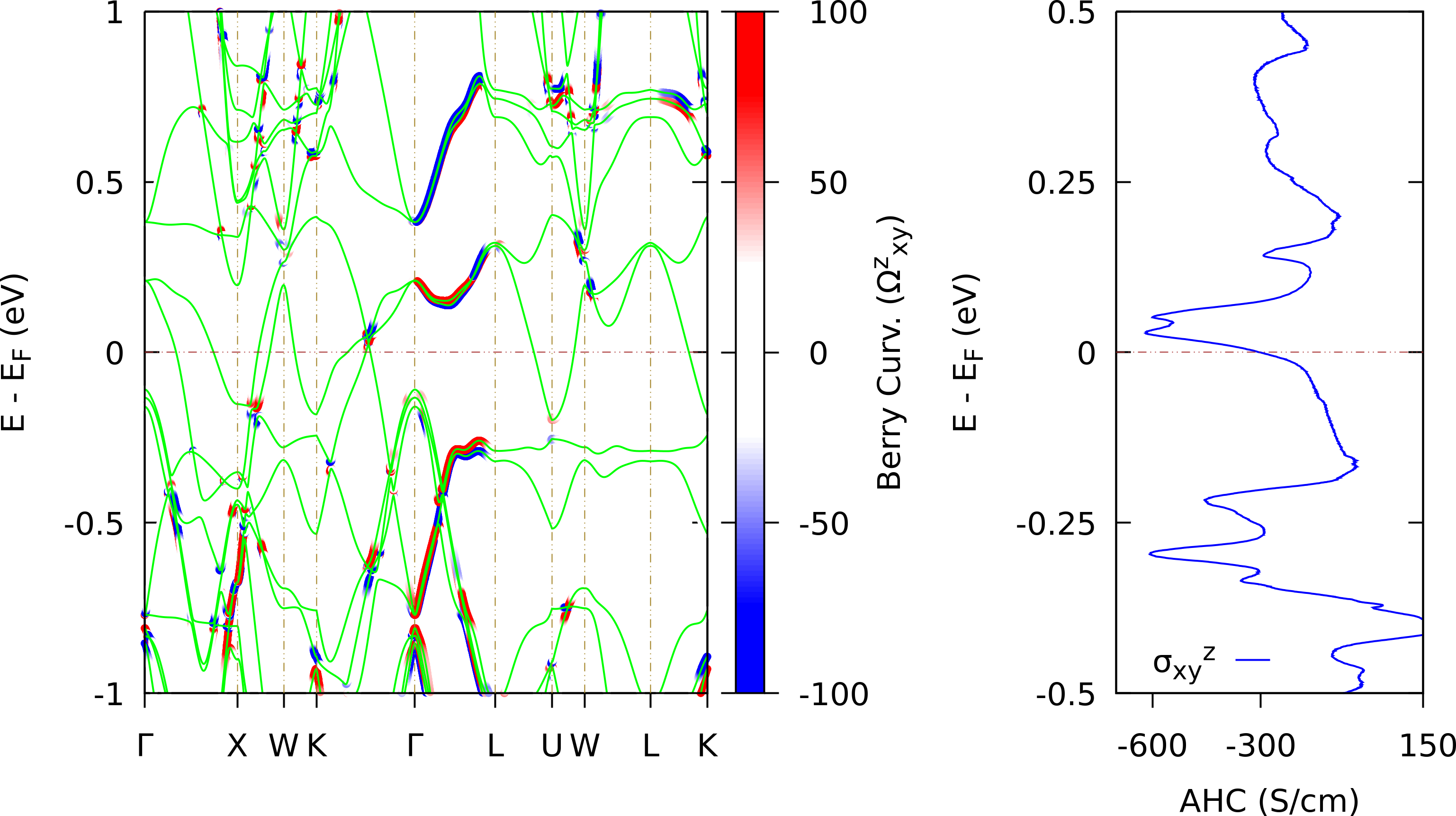}
\caption{\label{fig-band-structure} Left pannel: Band structure and Berry curvature (BC) contribution along high symmetry line of the cubic BZ in the GGA+SOC for ferro magnetic (FM) state with magnetic moment along the z-axis. Right pannel: Calculated anomalous Hall effect as a function of Fermi energy. }
\end{figure*}

An interesting aspect of the present results emerges from comparison with the parent Heusler compounds. The reported intrinsic AHC values of Co$_2$VAl, Co$_2$CrAl, and Co$_2$FeAl are approximately 44.7, 120, and $155~ \Omega^{-1}$.cm$^{-1}$, respectively, while Co$_2$TiAl is reported to exhibit predominantly extrinsic AHE with little or no intrinsic contribution \cite{Co2TiAl_intrinsic, Co2TiAl_extrinsic, Co2VAl_AHE_intrinsic, Co2VAl_1, Co2CrAl_AHE_temperaure_dependent, Co2CrAl_AHE_theory, Co2FeAl_shukla}. 
Considering the extensive chemical substitution and configurational disorder in \s~, the intrinsic Hall response might be expected to be significantly diluted, particularly since Co$_2$FeAl, which exhibits the largest intrinsic AHC among the parent compounds, contributes only 25\% of the transition-metal sublattice.
Contrary to this expectation, the entropy-stabilized system exhibits an anomalous Hall conductivity of $134.4~ \Omega^{-1}$.cm$^{-1}$, which is comparable to the largest values reported for the parent compounds and exceeds most of them. The persistence of such a high intrinsic AHC despite significant chemical disorder suggests that the Berry curvature driven phenomena remain robust in this system and cannot be understood as a simple averaging of the constituent compounds. Instead, the multi-element substitution appears to modify the electronic structure and Berry curvature distribution near the Fermi level, leading to an enhanced intrinsic Hall response. This behavior can be viewed as a manifestation of the cocktail effect, one of the characteristic core effects of entropy-stabilized materials, wherein the resulting physical properties cannot be predicted by a simple averaging of the constituent compounds. The present results therefore demonstrate that entropy engineering can be an effective route for tailoring Berry curvature driven transport phenomena while retaining a robust intrinsic anomalous Hall response even in the presence of substantial configurational disorder.
}

\section{Conclusions}
In conclusion, we have investigated the structural, magnetic, transport and magnetotransport properties of the entropy-stabilized Heusler alloy \s. It crystallizes in a cubic Heusler structure with signatures of partial chemical disorder and exhibits metallic transport together with soft ferromagnetic behavior. The experimentally obtained saturation magnetization is found to be close to the Slater--Pauling value, indicating that the essential electronic structure governing ferromagnetism remains largely intact despite substantial configurational disorder.
The temperature dependence of $\rho_{xx}$ follows metallic behavior and can be well described by the Bloch--Grüneisen model. The analysis indicates that the electrical transport is predominantly governed by electron--phonon scattering, while the magnon contribution remains negligible within temperature range of this study.
Hall measurements reveal anomalous Hall effect with the maximum anomalous Hall conductivity being $134.4~ \Omega^{-1}$.cm$^{-1}$ at 10 K. The anomalous Hall angle obtained in the present system is also relatively large. 
Combined experimental results and first-principles calculations establish that the anomalous Hall effect is predominantly intrinsic in origin and arises from the Berry curvature of the electronic bands.
Remarkably, despite extensive chemical substitution and significant configurational disorder, the anomalous Hall conductivity remains comparable to the highest values reported among the corresponding parent Heusler compounds. We attribute this behavior to be a manifestation of the cocktail effect of entropy-stabilized materials. The persistence of intrinsic anomalous Hall response further demonstrates the robustness of Berry-curvature-mediated transport against chemical disorder. Our findings suggest that entropy engineering may provide an effective strategy for tuning topological transport properties and establish entropy-stabilized Heusler alloys as promising platforms for realizing robust spintronic and Berry curvature driven properties.

\section{References}
\bibliographystyle{aipnum4-2}
\bibliography{bibliography.bib}

\end{document}